# Automatic Identification of Machine Learning-Specific Code Smells

Peter Hamfelt
Roxtec Group
Karlskrona, Sweden
peter.hamfelt@gmail.com

Ricardo Britto
Ericsson
Blekinge Institute of Technology
Karlskrona, Sweden
ricardo.britto@{ericsson.com,@bth.se}

Lincoln Rocha
Department of Computing
Federal University of Ceará
Fortaleza, Ceará, Brazil
lincoln@dc.ufc.br

Camilo Almendra
Department of Computing
Federal University of Ceará
Fortaleza, Ceará, Brazil
camilo.almendra@ufc.br

## ABSTRACT
Machine learning (ML) has rapidly grown in popularity, becoming vital to many industries. Currently, the research on code smells in ML applications lacks tools and studies that address the identification and validity of ML-specific code smells. This work investigates suitable methods and tools to design and develop a static code analysis tool (MLpylint) based on code smell criteria. This research employed the Design Science Methodology. In the problem identification phase, a literature review was conducted to identify ML-specific code smells. In solution design, a secondary literature review and consultations with experts were performed to select methods and tools for implementing the tool. We evaluated the tool on data from 160 open-source ML applications sourced from GitHub. We also conducted a static validation through an expert survey involving 15 ML professionals. The results indicate the effectiveness and usefulness of the MLpylint. We aim to extend our current approach by investigating ways to introduce MLpylint seamlessly into development workflows, fostering a more productive and innovative developer environment.

## 1 Introduction
Software quality has been widely studied [2, 12], but more studies have yet to be made on the software quality of machine learning (ML) applications. ML gains traction daily in today's fast-paced digital world, bringing additional complexity. Hence, it's essential to prioritize the development of sustainable and maintainable ML applications to minimize factors such as hidden technical debt [1, 21]. Even more so when empirical evidence shows that maintenance and technical debt management have unique challenges when developing ML software [15, 17, 22, 24].

Traditionally, static analysis tools have shown to be an effective way to identify potential code issues, such as code smells, early on in the development process and, as a result, improve software quality [13]. Although code smells have been thoroughly researched in traditional software systems, limited research has been done on their prevalence in ML systems [23]. Additionally, few studies have focused on ML application-specific code smells [27]. However, while some studies have proposed ML-specific code smells, very few have confirmed their validity or attempted to automate the process of statically detecting these smells. Moreover, there has been limited evaluation of such tools' practical relevance and applicability within the industry.

To address the specified research gap concerning the automated identification of ML-specific code smells, we developed and evaluated MLpylint, a stand-alone static code analysis tool tailored to detect ML-specific code smells in Python-based ML applications.

In this paper, we answered the following research questions:
**RQ1**. *What specific code smells are unique to Python machine learning (ML) applications?*
**RQ2**. *What is the most appropriate approach for implementing a static code analysis tool to detect ML-specific code smells?*
**RQ3**. *How well does the developed static analysis tool perform?*
**RQ4**. *How adequate is the developed tool in terms of its usefulness?*

This paper's main contributions are as follows:

- A review of best practices identified in previous research to develop a code smell criterion. This forms the basis for selecting the code smells to be incorporated into the MLpylint.
- A static code analysis tool to identify ML-specific code smells in Python-based ML applications.
- An empirical validation of the MLpylint.

The remainder of this paper includes a related work section, which reviews relevant related work and identifies gaps in the current research landscape. We described our research design and the developed solution, followed by a section describing the evaluation results. We concluded the paper by discussing the associated research validity threats, conclusions, and views on future work.

## 2 Related work
Code smells refer to specific patterns in the source code that suggest potential defects, indicate opportunities for refactoring, or highlight violations of common coding standards [4]. They do not necessarily signal errors or bugs but often point to areas in the code that can be problematic or less maintainable over time.

Literature on code smells unique to machine-learning applications is sparse, with only a handful of notable studies [3, 5, 23].

While these studies highlighted common code smells in standard software systems within the ML domain, none specifically focused on Python-based ML applications. However, an exception was a study by Zhang et al. which cataloged 22 code smells distinct to Python-based ML applications [27]. This research provided comprehensive details on each code smell, explaining its context, associated problem, and potential solution. Table 2 outlines a summary of the code smells identified by Zhang et al..

As the demand for ML applications increases, it is essential to prioritize maintainability and software quality to ensure their effectiveness and longevity. This includes addressing code smells that are specific to ML applications.

van Oort et al. [23] examined the frequency of common code smells in ML projects. Noting a general lack of software engineering know-how in AI and ML, they linked this deficit to decreased maintainability and reproducibility. They used Pylint, a Python-based static code analysis tool, to analyze 74 open-source Python ML projects for defects, refactoring needs, and regular coding standard violations. Their work spotlighted dependency management challenges in Python ML projects, emphasizing that dependency specifications are central to the projects' reproducibility and maintainability. Moreover, they pointed out Pylint's shortcomings in accurately checking import statements, especially for leading ML libraries like PyTorch. This problem makes continuous integration in ML systems more challenging, underscoring the need for further research in the field.

In the research conducted by Gulabovska and Porkolab [6] (2019), the shortcomings of abstract syntax tree (AST) based tools in the static analysis of Python are discussed. The study proposes integrating symbolic execution techniques with existing methods to better address Python's dynamic behaviors and improve vulnerability detection. Although this approach can potentially lead to false positives, the research recommends using AST-based and symbolic execution tools to maximize true positives while minimizing false alarms in Python static analysis.

Their works [6, 23] both point towards limitations in existing tools such as Pylint, which struggles with accurately checking import statements in leading ML libraries. The necessity for more robust tools that can mitigate such issues remains a critical concern.

Tang et al. (2021) explored the evolution and maintenance of ML systems, focusing on refactorings and technical debt issues specific to ML [25]. Analyzing 26 open-source ML projects containing 4.2 million lines of code and 327 scanned code patches, they shed light on why developers undertake refactorings, such as addressing technical debt, avoiding code duplication, and solving ML-specific problems. The team introduced 14 new ML-specific refactorings and pinpointed seven new categories of ML-specific technical debt, offering insights that can aid practitioners, tool creators, and educators in enhancing the longevity and efficiency of ML systems by understanding and managing technical debt better.

Alahdab [1] (2019) explored the emergence of Hidden Technical Debt (HTD) patterns during the early development, particularly the prototyping stage, of ML applications through a case study involving the public transportation agency Vasttrafik in Sweden. The study identified 12 out of 25 potential HTD patterns mainly involving underutilized data dependencies and ML code smells, emphasizing the need to address these issues early in the development stages. The results suggested a broader relevance of the findings across various ML application domains, underscoring the need for refinement before reaching the production phase to maintain software engineering standards. The study indicates a need for strategies that not just identify but also resolve these issues at an early stage, possibly through automated tools.

Lacerda et al. (2020) [13] examined the relationship between "code smells" and refactoring through a detailed literature review spanning 1992 to 2018. The study identified a significant link between code smells and the refactoring process affecting software quality. It underscored the diversity and inconsistency in the current code-smell detection tools and strategies, pointing out a shift towards probabilistic or search-based tools despite the prevalent use of metric-based tools. The research calls attention to the emergent potential of recent tools while stressing the necessity for more uniform approaches to code-smell detection.

Ichtsis et al. (2021) points out the difficulties in identifying code smells accurately, which is a key part of understanding Technical Debt (TD) [16]. These "smells" are found using tools that scan the code for specific issues, representing areas where the coding could be improved. Although these detections are crucial in TD analyses, the process is not straightforward and can yield unclear results. Even when using a specialized tool that combines different detection methods, there was a noticeable lack of agreement on what comprises a "code smell."

Paiva et al. (2017) highlighted the complexity of comparing code smell detection tools due to the informal and varying definitions of code smells used by different researchers and developers [19]. This discrepancy leads to tools having distinct detection techniques and results, impacting the validation time. The study underscores the need for a standardized approach to defining and identifying code smells to enhance the effectiveness and efficiency of these tools.

Both the studies by Lacerda et al. [13] and the collaborations of Paiva et al.. [19] and Ichtsis et al. [16] underline the pressing issue of inconsistency in the current "code smell" definitions and the strategies employed to detect them. These differences lead to varying effectiveness and reliability of different detection tools, presenting a clear pathway for future work to standardize definitions, criteria, and detection strategies related to code smells.

In conclusion, the discussed related works highlight the importance of software engineering principles and practices in developing and maintaining ML systems. These studies provide insights into the specific challenges and issues that arise in the field of ML, such as code smells and technical debt and offer potential solutions, such as automated static analysis tools, to address them. Yet, there remain enduring challenges, hinted in the reviewed studies, showcasing a need for continued exploration and progress in this area.

## 3 Research Design

To answer our research questions, we employed the Design Science Methodology (DSM), following the design process proposed by Offermann et al. [18]. DSM operates in three critical stages: problem identification, solution design, and evaluation. The investigation was conducted at Ericsson[1], a global telecommunication vendor.

---
[1]Ericsson. https://www.ericsson.com

We provide more details of our research design in the remainder of this section, per phase of the design science process.

### 3.1 Problem Identification

In the problem identification phase, Ericsson highlighted the crucial need to investigate ML-specific code smells in machine-learning applications. The goal was to reduce technical debt and promote the development of more maintainable software.

To address RQ1, we conducted a comprehensive literature review. The details of this review, including search engines and keywords employed, can be found in [8]. Despite rigorous exploration, it was found that literature on code smells unique to machine-learning applications is sparse, with only a handful of notable studies [3, 5, 23]. While these studies highlighted common code smells in standard software systems within the ML domain, none specifically focused on Python-based ML applications.

To address RQ2, we investigated the best practices of code smell software engineering to establish the criteria that render code smells statically analyzable. We also sought expert insights from the Python Code Quality Authority (PyCQA). The PyCQA [2] is an informal collective of maintainers behind widely used Python tools for automated style and quality checks. It exists mainly to give these related projects a single, easy-to-find GitHub namespace and to foster collaboration among their volunteer developers. As a result, we decided to use the open-source tool Pylint[3] as a basis for MLpylint. Pylint uses the Python library Astroid [4], which provides the functionality for parsing the abstract syntax tree (AST) and inferring objects.

We interacted with the PyCQA maintainer community to identify the best way to answer our research questions. Based on the nature of the identified ML-specific code smells, most of them are linked to libraries (e.g., PyTorch and TensorFlow) that are not part of the standard Python module library. The PyCQA maintainers stated that Pylint could not infer external modules and recommended not extending it. Instead, the recommendation was to develop a new tool to leverage Astroid's capabilities fully, which led to the MLpylint's development.

Based on our literature review and interactions with the Pylint/Astroid maintainer community, we specified the code smell criteria implemented in MLpylint (see Table 1).

| ID | Criterion |
|---|---|
| CSC1 | **Smell Descriptions:** The code smell includes context, problem, and solution description. |
| CSC2 | **Domain Relevance:** The code smell is specifically relevant to machine learning within Python. |
| CSC3 | **Static Detectability:** The code smell can be detected through static analysis without code execution. |
| CSC4 | **Distinct Pattern:** The code smell pattern is identifiable in Python source code. |
| CSC5 | **Precision of Pattern:** The code smell exhibits a clear and unambiguous pattern. *Determines code smell classification.* |

**Table 1: Code smell criteria implemented in the MLpylint.**

---

[2]PyCQA. https://meta.pycqa.org/introduction.html
[3]Pylint. https://pylint.readthedocs.io/en/latest/
[4]Astroid. https://pylint.readthedocs.io/projects/astroid/en/latest/index.html

Each identified code smell undergoes assessment based on specific predefined criteria. After this assessment, the code smells fall into one of the following classifications: **Code Smell (CS)**: These are clear patterns visible directly in the code, with little to no ambiguity associated with the code smell identification pattern; **Code Smell Advice (CSA)**: These suggest potential issues or inefficiencies linked to a specific code smell pattern. However, this pattern is often too general to be definitive by itself. Consequently, a thorough analysis by the developer is required to determine if it is indeed associated with the particular code smell; and **Not Applicable (NA)**: This category includes any code smell that doesn't meet the established criteria.

After careful review, we deemed as relevant 20 of the 22 code smells identified by Zhang *et al.* [27]. Of these, 14 were labeled as code smells, and 6 as code smell advice. Details on the classification against the code smell criteria can be found in Table 2.

### 3.2 Solution Design
### 3.3 Evaluation

The evaluation focused on assessing MLpylint's performance, feasibility, and usefulness. We selected 160 open-source Python ML application repositories from GitHub to evaluate their performance. The selection process was primarily driven by the characteristics of code smells identified in the problem identification phase.

Five search strings derived from these code smell characteristics were employed to pinpoint related projects specific to Python ML applications. Additionally, a broader search string was used to encompass general themes in machine learning, artificial intelligence, and data science. In the following, we list each search query string.

- Pandas: machine learning in:readme in:description AND pandas in:file language:python pushed:>2023-01-01
- Scikit-Learn: machine learning in:readme in:description AND sklearn in:file language:python pushed:>2023-01-01
- Pytorch: machine learning in:readme in:description AND torch in:file language:python pushed:>2023-01-01
- NumPy: machine learning in:readme in:description AND numpy in:file language:python pushed:>2023-01-01
- Tensorflow: machine learning in:readme in:description AND tensor in:file language:python pushed:>2023-01-01
- ML/AI/DS: "machine learning" OR "artificial intelligence" OR "data science" in:readme in:description language:python pushed:>2023-01-01

To ensure the reproducibility of this research and to capture a consistent version of each project during analysis, every repository was forked. This action guaranteed the preservation of each project's exact state and version at the time of the study. The complete list of these identified projects is available in [11]. Each selected repository was carefully reviewed by at least two authors to confirm its relevance and adequacy to the study goals. The criteria used for this evaluation follows: **C1**: The project should be explicitly centered on "Machine learning"; **C2**: The codebase should be written in the Python language; **C3**: There must have been commits (or "pushes") to the project after January 1, 2023; **C4**: The repository should not be a tutorial or guide but should represent an actual application; and **C5**: On GitHub, the "sort by" option needs to be set to "Best match".

| Code Smell (from Zhang *et al.*) | CSC1 | CSC2 | CSC3 | CSC4 | CSC5 | Classification in MLpylint |
|---|---|---|---|---|---|---|
| 1 Unnecessary Iteration | ✓ | ✓ | ✓ | ✓ | ✗ | CSA |
| 2 NaN Equivalence Comparison | ✓ | ✓ | ✓ | ✓ | ✓ | CS |
| 3 Chain Indexing | ✓ | ✓ | ✓ | ✓ | ✓ | CS |
| 4 Columns and DataType Not Explicitly Set | ✓ | ✓ | ✓ | ✓ | ✓ | CS |
| 5 Empty Column Misinitialization | ✓ | ✓ | ✓ | ✓ | ✓ | CS |
| 6 Merge API Parameter Not Explicitly Set | ✓ | ✓ | ✓ | ✓ | ✓ | CS |
| 7 In-Place APIs Misused | ✓ | ✓ | ✓ | ✓ | ✓ | CS |
| 8 Dataframe Conversion API Misused | ✓ | ✓ | ✓ | ✓ | ✓ | CS |
| 9 Matrix Multiplication API Misused | ✓ | ✓ | ✓ | ✓ | ✓ | CS |
| 10 No Scaling before Scaling-Sensitive Operation | ✓ | ✓ | ✗ | ✗ | - | NA |
| 11 Hyperparameter Not Explicitly Set | ✓ | ✓ | ✓ | ✓ | ✓ | CS |
| 12 Memory Not Freed | ✓ | ✓ | ✓ | ✓ | ✗ | CSA |
| 13 Deterministic Algorithm Option Not Used | ✓ | ✓ | ✓ | ✓ | ✓ | CS |
| 14 Randomness Uncontrolled | ✓ | ✓ | ✓ | ✓ | ✓ | CS |
| 15 Missing the Mask of Invalid Value | ✓ | ✓ | ✓ | ✓ | ✓ | CS |
| 16 Broadcasting Feature Not Used | ✓ | ✓ | ✓ | ✓ | ✗ | CSA |
| 17 TensorArray Not Used | ✓ | ✓ | ✓ | ✓ | ✓ | CS |
| 18 Training / Evaluation Mode Improper Toggling | ✓ | ✓ | ✓ | ✓ | ✗ | CSA |
| 19 Pytorch Call Method Misused | ✓ | ✓ | ✓ | ✓ | ✓ | CS |
| 20 Gradients Not Cleared before Backward Propagation | ✓ | ✓ | ✓ | ✓ | ✗ | CSA |
| 21 Data Leakage | ✓ | ✓ | ✓ | ✓ | ✗ | CSA |
| 22 Threshold-Dependent Validation | ✓ | ✓ | ✗ | ✗ | - | NA |

Table 2: Result matrix of code smells meeting criteria

To address RQ3, we employed proportional stratified random sampling and manual validation to check the accuracy of our tool, discussed in Section 5.1.

To address RQ4, we performed a qualitative study with practitioners, discussed in Section 5.2. We conveniently sampled 15 developers from Ericsson who actively worked in the field of ML. Participants received an initial demonstration detailing the tool's installation process and showcasing all its features. Then, participants were given a month to utilize the tool within their own ML application environments to gain firsthand experience.

## 4 MLpylint overview

MLpylint is a static code analysis tool designed to automate the identification and evaluation of code smells specific to machine learning in Python source code. The tool is designed to aid the development of ML applications by supporting developers, data scientists, ML engineers, and data engineers, particularly those not well-acquainted with traditional software engineering practices, empowering them to contribute effectively [1]. This section delves into key design choices, the architecture and functionality of the tool, shedding light on its key modules and how they contribute to its capabilities (for more details on tool's usage and instructions refer to PyPI repo [7], for source code refer to GitHub repo [9]).

### 4.1 Architecture

Figure 1 provides an architectural overview. At the developer's workstation, the tool resides within the ML-Application Project and is incorporated into the Python virtual environment. MLpylint offers manual checks on Python source code. Additionally, it can be seamlessly activated by a pre-commit hook during git commit operations. Following the commitment and push of changes to a GitLab repository, a merge request is generated, which subsequently activates the CI/CD pipeline. While this pipeline includes routine tasks like compilation and integration tests, it distinctly embeds the analysis phase. The orange highlights serve to underline the pivotal role of MLpylint, illustrating its significance both in the developer's workflow and the CI/CD process. For ease of access and integration, MLpylint is readily procurable from the PyPi Repository [5].

The tool architecture comprises three modules and utilizes a third-party library. The Analysis Runner initiates the tool based on input parameters and manages the Python file extraction process. It utilizes Astroid, a Python library, to parse Python files and extract the abstract syntax tree. Code Smell Checkers is responsible for pinpointing code smells. The Results module accumulates and arranges the findings. Finally, the Analysis Report prepares an output of the analysis findings for the developers.

As illustrated in Figure 2, the workflow initiates with the analysis runner processing the input parameters defined by the developer. It methodically iterates over Python files nested within sub-directories based on the input path. Upon identifying a .py file, the analysis runner directs it to Astroid to extract its abstract syntax tree. The sub-sequence section explains Astroid and how the parsing of the abstract syntax tree is accomplished. Furthermore, a short explanation and example of the AST representation will be demonstrated.

Astroid is a Python library package for abstract syntax tree (AST) parsing and static code analysis. It also has functionality for inferring the value of objects. In MLpylint, Astroid is used to convert Python code into its AST representation. Beyond that, it aids in inferring values, which involves predicting potential outcomes during the run-time execution of the Python code. It's important to highlight that Astroid cannot infer values from external sources not included in the parsed Python file; managing such external dependencies would necessitate extra functionality.

---
[5]MLpylint. https://pypi.org/project/mlpylint/

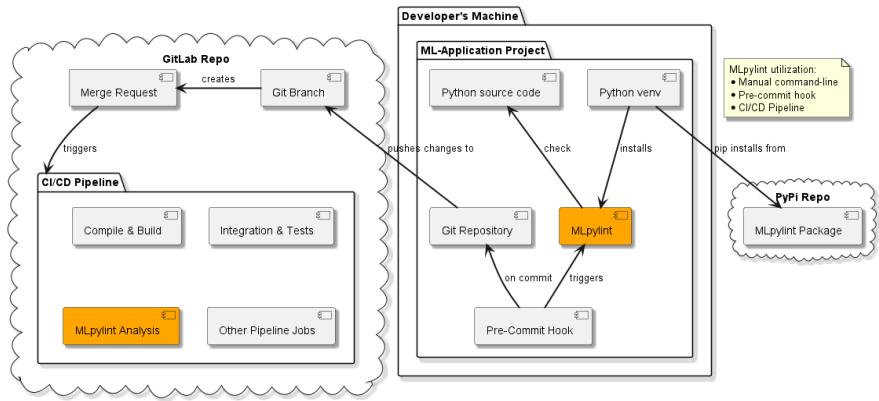

Figure 1: Architectural overview showing how MLpylint integrates into a typical ML-application project code and CI/CD.

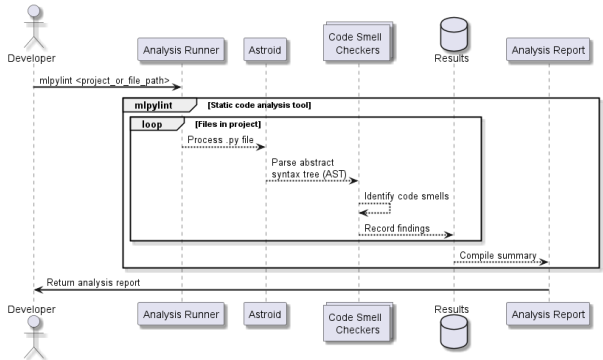

Figure 2: Sequence diagram showing the internal flow and components of MLpylint.

At the heart of MLpylint's functionality lie the code smell checkers, which drive its primary operations. These checkers have been methodically designed, adhering to a specific hierarchy and structure:

(1) Base Checker Class - The base checker class is at the foundational level. Every specific code smell checker is derived from this foundational class.
(2) Recursive Node Iteration - Integral to the base checker class is its capability to iterate recursively through every node present within an AST module. This iteration ensures extensive analysis.
(3) Visitor Pattern - This recursive iteration manifests the visitor pattern. This design pattern is particularly effective for tasks like AST-based code analysis because it allows for a modular and scalable approach.
(4) Custom Node Visitation - By employing the visitor pattern, each code smell checker can specify which nodes they are interested in. This means that every node does not need to be analyzed; they can target specific nodes relevant to their code smell pattern.
(5) Dedicated Logic Implementation - Once the relevant nodes are identified, each checker can apply its unique logic and functionality within the scope of its dedicated class instance. This encapsulation ensures that each code smell has its distinct processing logic, making the system modular and maintainable.

MLpylint's design ensures thoroughness in code analysis and clarity in its implementation, with each code smell checker working autonomously yet cohesively within the broader framework. Code smells are classified into two distinct categories, each serving a unique purpose: Code Smells and Code Smells Advice.

**Code Smells:** These are clear, unambiguous patterns identified directly in the code. Their detection is precise, and there's little to no room for ambiguity regarding their presence. For instance, a code smell might arise if a particular function doesn't set all the required keywords when invoked. In such a case, detecting the function call and validating all the keywords are set is possible.

**Code Smells Advice:** This category captures potential issues that are broader in nature. They hint at possible problems or inefficiencies but are not definitive. Detecting these requires deeper analysis, often necessitating a developer's intervention to ascertain if there's an actual code smell. For example, potential memory leakage caused by a specific ML model could be flagged under this category. While the initial indicator might suggest a problem, a thorough investigation is required to confirm the issue.

To summarize, while Code Smells provide specific and definitive instances of sub-optimal code patterns, Code Smells Advice points to areas that might need closer scrutiny, leaving the final investigation up to the developer. The Results module manages and stores the identified code smells, ensuring that relevant details of each detection are stored. Upon completion of the analysis, these detected code smells are effectively relayed to the developer via the analysis report module.

# 5 Results

In this section, we present the results associated with MLpylint's evaluation.

## 5.1 Performance Evaluation

In addressing RQ3, the tool performance evaluation results are presented, highlighting its ability to identify ML-specific code smells in Python source code files. The primary objective of this evaluation is to closely examine the data generated by the tool, assessing its precision in pinpointing code anomalies.

During the evaluation, MLpylint was utilized to analyze 160 open-source ML application projects available on GitHub, concentrating on its performance and effectiveness in detecting code smells. Detailed information regarding the project selection is documented in Section 3.3. The data gathered during this analysis, summarized in Table 3, provides crucial insights into the working dynamics of the tool and its proficiency in recognizing code smells. Observations from the results indicate that it was capable of processing 15 GB of project data, consisting of 86,577 files housed in 16,806 folders, in a duration slightly below 36 minutes, showcasing its efficient performance speed.

A substantial focus was examining Python files, with 33,360 .py files analyzed. This culminated in a total line count of 10,015,028 lines of Python code (LOC), yielding an average of approximately 300 lines per .py file. It is important to consider the nature of the analyzed projects, as more mature projects naturally house more extensive scripts than smaller, newer projects. Furthermore, including longer complex data processing scripts and algorithms, common in ML applications, can significantly increase the line count.

Regarding the Python files inspected, MLpylint identified code smells in 3,877 files, equating to approximately 11.6% of the total Python files examined. In addition, it is notable that 807 files encountered parsing difficulties due to syntax errors, emphasizing the necessity for clean and error-free coding practices. This substantial data pool offers a rich ground for revealing patterns and drawing significant conclusions regarding code quality in ML applications, supporting focused efforts towards improved coding practices in the field.

Before the discussion of code smells, we discuss the analysis coverage metrics, summarized in Table 3. It showcases a low syntax error rate of 2.4%, implying syntax errors' prevalence in open-source projects on GitHub. The code coverage is another critical aspect that has been measured. The results show that only 37.6% of the analyzed files were Python (.py). While this metric indicates that a substantial portion of the projects were composed of other file types, it is not necessarily a limitation, given that ML application projects can naturally include various file elements. In future analyses, exploring the roles and impacts of these other file types in ML application projects could be beneficial in providing a more comprehensive view of the project structures and dynamics.

We also evaluated MLpylint's performance from a speed perspective. It completed the analysis of all 160 projects within 35 minutes and 54 seconds. This translates to an average runtime per project of just 13.4 seconds, showcasing the tool's efficiency and speed in analyzing projects.

Regarding the code smell identification, the Tables 4 and 5 summarize the prevalence of smells in the analyzed projects.

The analysis uncovered a significant prevalence of specific code smells, with CS14 being the most frequently occurring, recorded 1796 times. On the code smell advice front, CSA12 and CSA21 have over a thousand instances each, suggesting they are the primary code smell advice the tool resorts to in addressing code smells. Furthermore, some code smells, such as CS5, CS9, CS13, and CS19, were not detected in the analyzed dataset, thereby highlighting a non-occurrence of these issues in the examined Projects.

Looking at the code smell advice section, a notable number of instances are seen where the tool has generated advice, indicative of the potential areas that recommend further investigation by developers. The gathered data not only presents an insight into the prevailing code smells but also brings to light the specific areas developers are being advised to focus on.

To further analyze the findings, proportional stratified random sampling was carried out separately for both the code smell advice and smell instances. This strategy aims to reduce the sampling error and ensure a precise analysis representation of each stratum [14, 26]. Table 6 displays the breakdown and validation results of the identified code smell advice instances. It lists the total number of occurrences for each code smell advice ID, size in terms of the number of occurrences, the proportion they represent of the total in percentage, and the derived sample size from the stratified random sampling method. Additionally, it portrays the counts of true and false positives.

| Data Collection | Results |
| --- | --- |
| Run-time of analyzing all projects | 35m 54s |
| Number of projects analyzed | 160 |
| Size of all projects (GB) | 15 GB |
| Lines of Python code | 10,015,028 |
| Number of code smells identified | 5,380 |
| Number of folders identified | 16,806 |
| Number of files identified | 86,577 |
| Number of .py files analyzed | 33,360 |
| Number of .py files with detected code smells | 3,877 |
| Number of .py files unable to parse due to syntax error | 807 |
| **Metrics** | **Results** |
| Rate of Syntax Errors | 2.4% |
| Code Coverage | 37.6% |

Table 3: Summary of data points gathered from the static analysis and coverage metrics

| Code Smell | Instances |
| --- | --- |
| CS2 | 15 |
| CS3 | 33 |
| CS4 | 893 |
| CS5 | 0 |
| CS6 | 13 |
| CS7 | 14 |
| CS8 | 32 |
| CS9 | 0 |
| CS11 | 26 |
| CS13 | 0 |
| CS14 | 1796 |
| CS15 | 48 |
| CS17 | 4 |
| CS19 | 0 |
| Total | 2874 |

Table 4: Code Smell instances identified

| Code Smell Advice | Instances |
| --- | --- |
| CSA1 | 87 |
| CSA12 | 1041 |
| CSA16 | 38 |
| CSA18 | 13 |
| CSA20 | 285 |
| CSA21 | 1042 |
| Total | 2506 |

Table 5: Code Smell Advice instances identified

Looking at the data, it is clear that CSA12 and CSA21 are the most common instances. CSA12 has almost an equal number of true and false positives, indicating that it might sometimes give incorrect advice. On the other hand, CSA21 has a high number of true positives, showing that it generally gives correct smell detection.

During the manual validation, specific observations were made that brought attention to several potential areas for enhancement. A notable observation under CSA1 was the incorrect inference spotted in different scenarios in various configurations and in the iteration over data frame columns.

Additionally, it was observed that CSA12 tended to be too broad in its smell detection, often flagging correct implementations as smells, especially in contexts involving memory management and loop structures involving TensorFlow or Torch objects. The CSA20 smell exhibited incorrect inference and string detection issues in several code expressions and function calls, indicating a need for refinement in its detection mechanisms.

CSA21 flagged certain instances as false positives by identifying a method used in one library (transformers pipeline) as a code smell when it was related to a different library (sklearn pipeline). This demonstrates that the code smell pattern mechanism of this advice needs further refinement to distinguish between methods from different libraries accurately.

Table 7 outlines the distribution and validation of results for the code smells identified. The results unveil a consistent trend of true positives across all sampled code smells, with no obvious false positives noted.

During the manual validation process, several observations provided insight into the potential pitfalls and areas of concern in analyzing the ML application projects. For both CS4 and CS11, it was observed that parameters might sometimes be set without using keywords to indicate the specific parameter being set, which tends to decrease readability. This is a clear signal for developers to maintain a practice of explicitly specifying keywords for parameters to foster more readable and maintainable code.

A particularly recurrent code smell, CS14, highlighted a complex issue regarding using randomness seeds in code. On the one hand, it was noted that while seeds were being set, they were not globally applied, raising questions regarding the desirability of this approach, especially in the context where isolation of randomness might be desirable or not. Furthermore, the analysis highlighted concerns regarding seeds being imported from other modules without checks, potentially leading to undesirable code smells. It was noted that a single seed was often used for only one package or module despite multiple packages or modules in the same file that did not have their specified seed set. This brings up a crucial question about the intention behind such configurations. As a result, it is up to the developer to determine if this constitutes a code smell in their particular scenario, urging a thoughtful evaluation to uphold the standard of code quality.

This outcome shows that the tool can reliably identify code smells. CS14 was the most common issue in the derived data, appearing much more frequently than others. This suggests that it is a common issue and needs more attention for enhanced code quality.

Table 8 provides a detailed view of tool's ability to identify code smells. It is observed that the tool displays a high precision rate in detecting Code Smells (CS) with a score of 100%. However, it has room for improvement in identifying Code Smell Advice (CSA), where it scored 73.6%, resulting in an overall precision of 87.9%. The results depict a fairly balanced code smell distribution between CS and CSA, presenting an almost equal propensity for both types. Lastly, the density of code smells, calculated per thousand lines of code (KLOC), is found to be 0.53, measuring the concentration of code smells present in the analyzed projects.

## 5.2 Feasibility and User Evaluation

To address RQ4, we obtained the feedback of 15 practitioners via a survey. Regarding their organizational roles of the participants, 14 (93%) participants identify as Data Scientists, while one (7%) is positioned as ML Architect. As for their years of experience in machine learning development, three (20%) participants have 1 to 5 years of experience, seven (47%) have between 5 to 10 years, and five (33%) boast more than 10 years of experience. This suggests that over half of the respondents possess over 5 years of experience in machine learning development, concluding that the majority are seasoned professionals in the field.

| ID | Size | Perc (%) | Sample | TP | FP |
|---|---|---|---|---|---|
| CSA1 | 87 | 3.5 | 12 | 10 | 2 |
| CSA12 | 1041 | 41.4 | 143 | 65 | 78 |
| CSA16 | 38 | 1.5 | 5 | 5 | 0 |
| CSA18 | 13 | 0.5 | 2 | 2 | 0 |
| CSA20 | 285 | 11.4 | 39 | 34 | 5 |
| CSA21 | 1042 | 41.7 | 144 | 138 | 6 |
| Total | 2506 | 100 | 345 | 254 | 91 |

Table 6: Proportional stratified random sampling and validation results for Code Smell Advice (T/FP = True/False Positive)

| ID | Size | Perc (%) | Sample | TP | FP |
|---|---|---|---|---|---|
| CS2 | 15 | 0.5 | 1 | 1 | 0 |
| CS3 | 33 | 1.1 | 4 | 4 | 0 |
| CS4 | 893 | 31.1 | 109 | 109 | 0 |
| CS5 | 0 | 0 | 0 | 0 | 0 |
| CS6 | 13 | 0.5 | 2 | 2 | 0 |
| CS7 | 14 | 0.5 | 2 | 2 | 0 |
| CS8 | 32 | 1.1 | 4 | 4 | 0 |
| CS9 | 0 | 0 | 0 | 0 | 0 |
| CS11 | 26 | 1 | 3 | 3 | 0 |
| CS13 | 0 | 0 | 0 | 0 | 0 |
| CS14 | 1796 | 62.5 | 219 | 219 | 0 |
| CS15 | 48 | 1.7 | 6 | 6 | 0 |
| CS17 | 4 | 0.1 | 1 | 1 | 0 |
| CS19 | 0 | 0 | 0 | 0 | 0 |
| Total | 2874 | 100 | 351 | 351 | 0 |

Table 7: Proportional stratified random sampling and validation results for Code Smells (T/FP = True/False Positive)

| Metric | Result | | |
|---|---|---|---|
| | CS | CSA | Total |
| Average code smells per project | 17.96 | 15.66 | 33.63 |
| Precision | 100% | 73.6% | 87.9% |
| Code Smell Distribution | 53.4% | 46.6% | – |
| Density of Code Smells (per KLOC) | 0.2869 | 0.2502 | 0.5371 |

Table 8: Code smells detection results

Figure 3 presents all the participants' feedback on six questions. Regarding the usability and ease of understanding (Q1), the results indicate that over 33% strongly agree that the tool is user-friendly and intuitive, while an additional close to 67% agree with this sentiment. Overall, this suggests that the respondents find it easy to use and understand.

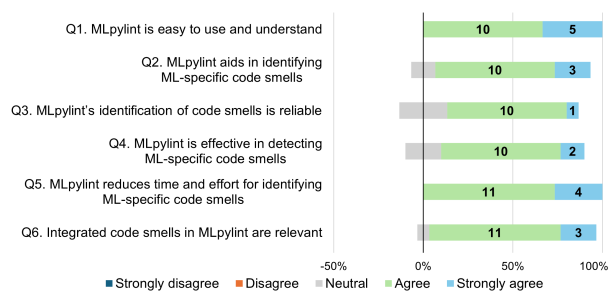

Figure 3: Participants' feedback on quality aspects

Figure 3–Q2 presents the participants' feedback regarding tool's effectiveness in aiding developers to identify and resolve code smells unique to machine learning within their Python-based ML applications. According to the data, 20% strongly agree that it is helpful, approximately 67% agree, and just over 13% remain neutral. These results suggest that a clear majority of participants believe it assists developers in pinpointing code smells in Python-based ML applications.

Figure 3–Q3 displays participants' responses to the reliability of tool's automatic identification of ML-specific code smells in Python-based ML applications. The data suggests that approximately 7% strongly agree with its reliability, around 67% agree, while nearly 27% remain neutral. From these findings, we observed that the majority of participants consider tool's automatic detection of ML-specific code smells to be dependable.

Figure 3–Q4 presents the responses to how effective the tool detects ML application-specific code smells within Python-based source code. The findings indicate that over 13% of participants strongly agreed, nearly 67% agreed, and 20% remained neutral. This suggests that most participants believe it is effective in identifying code smells in ML applications.

Figure 3–Q5 presents the responses to the question of whether the tool aids in reducing the time and effort required to identify ML-specific code smells within ML application projects. The findings indicate that approximately 27% of participants strongly agreed, while an overwhelming 73% agreed. This underscores that the majority of respondents believe it assists developers in reducing time and effort in identifying ML-specific code smells in ML applications.

Figure 3–Q6 presents the responses concerning the relevance of code smells integrated into the tool and whether they are frequently encountered during machine learning development. The data reveals that 20% of participants strongly agree, approximately 73% agree, and nearly 7% remain neutral. These findings suggest that a significant majority of participants concur that the code smells incorporated in it are both relevant and commonly encountered in their daily ML development tasks.

Table 9 showcases feedback on functionality, accuracy, and usability. While 35% of participants didn't provide feedback, 67% shared their experiences. The feedback provided highlighted its utility in performing basic checks but emphasized the ongoing need for expert reviews. One of the suggestions was to incorporate its capabilities into existing tools like lint, and hopes for expanded support for additional ML libraries, particularly within CI/CD pipelines. Concerns about dependency management surfaced, with users suggesting a more flexible tool configuration. Recommendations for future upgrades included adding support for Pyspark and enhancing TensorFlow-specific checks. While many users appreciated the current capabilities of MLpylint, they also expressed a desire for continuous updates based on evolving ML practices and regular user feedback. Eleven (out of 15) participants expressed interest in using the tool and providing feedback following the evaluation.

| Answers |
|---|
| Some hygiene checks can be done via this tool, but a specialist review is still critical. |
| Making it part of a well-known tool like lint can help, although they may want to remain specific libraries agnostic. |
| Would be nice to see further ml library inclusions and also have it integrated into CI/CD pipelines. |
| The package seems to install several dependencies, in cases where production environments may constrain some packages or if only specific python versions may be supported, a configurable interface may be helpful (if feasible). Other than this, I used the package for a few cases, and the results were satisfactory. I would try to use it further in the projects I am working on. Very good work! |
| Spreading a few typical use case tutorials would be interesting. |
| So far, good with the existing stuff. |
| This could be extended for Pyspark in the future. For Tensorflow code use of tf.functions for large code can be another check for code smell. |
| It is a great tool. As more ML code-smells are devised, MLpylint should accommodate them. |
| Maybe more MLpylint usability feedback data (in actual code where it is analyzed) needs to be collected from Python users and used to improve the tool further. |
| A library has always room for improvement. For example, the time complexities can be improved, new smells can be added for future scopes, etc. |

Table 9: Participants' feedback on the functionality and usability

## 6 Discussion

**Code Smell in Python ML Applications.** When examining the identified code smells closely during this research, it becomes apparent that most of them are associated with external Python libraries, implying that these code smells are not part of Python's built-in module. As a result, one must explicitly import a library to encounter the related code smell. Indeed, prior studies [6, 23] found that existing Python static code analysis tools like Pylint cannot verify the correct usage of imported dependencies, including well-known ML libraries like PyTorch. These findings indicate that the focus of technical debt concerning ML-based applications has partially shifted from in-house code to third-party library usage.

MLpylint can infer external modules during static analysis, requiring the installation of necessary external libraries in the Python virtual environment. This installation is critical as a majority of code smells are connected to external libraries such as Pandas, NumPy, scikit-learn, PyTorch, and TensorFlow. Although this makes it resource-intensive, it is essential for a detailed analysis. Future iterations should focus on streamlining the installation process and enhancing configurability to analyze selected libraries exclusively, thus reducing resource allocation.

**Code Smell Criteria.** Despite the recognized concept of "code smell", there are no universal criteria for static analyzability [13, 16, 19]. Different tools define and incorporate code smells in various ways, leading to differing outcomes for the same code smell. Recognizing this, MLpylint incorporates general rules and criteria that are partially based on the characteristics of identified ML-specific code smells, emphasizing their relevance to machine learning and that they include clear context, problem, and solution descriptions with distinct identification patterns. It became apparent that the code smells varied notably in terms of their specificity and scope, leading to a distinction between Code Smell and Code Smell Advice.

Zhang *et al.* [27] catalogued 22 ML-specific code smells. With the introduction of dslinter, an add-on plugin to Pylint, the initial steps toward validating these identified smells were made. In our study, we applied the code smell criteria to this catalogue, finding that 20 of the 22 smells are relevant (14 categorized as Code Smell and six as Code Smell Advice). The excluded ones lacked a detailed pattern for identification, while the ones tagged as advice were broader in their descriptions, sometimes indicating a general concept appearing across various levels, which necessitated a multi-indicator approach for detection. Looking forward, it is important to encourage more standardized static code analysis tools through deeper research into this field. We believe that a promising agenda for the community could be the development of a hierarchy of smell abstractions, so that tool builders could compose complex smells as a set of finer-grained and context-driven checks.

**Performance.** The tool managed to process a considerable volume of data in about 35 minutes, thus demonstrating a capability to handle large projects effectively. However, it faced difficulties analyzing 807 files due to syntax errors, highlighting a common issue in open-source projects on GitHub. The appearance of syntax errors suggests the importance of promoting better coding practices and the adoption of static code analysis in ML application projects.

The analysis revealed that the most frequently detected code smell was CS14 (1796 occurrences), suggesting a considerable gap in controlling randomness in ML application codebases. The prevalent use of randomness might be a standard practice in ML applications due to the nature of the data utilized. However, it poses a question regarding the need to regulate randomness to ensure reproducibility, indicating an area that requires attention from developers.

MLpylint mainly identified the code smell advices CSA12 and CSA21. While it provides valuable advice, there are instances when it might not fully align with the specific context, given the broad spectrum of the CSA pattern it covers. Manual analysis was guided by proportional stratified random sampling, revealing a consistent true positive rate for Code Smell detection. This suggests a reliable detection capability. However, there were occurrences of false positives in the Code Smell Advice, indicating a potential for refining the advisory mechanism to prevent incorrect code smell warnings.

**User Evaluation.** In terms of MLpylint's usability, as depicted in Figure 3–Q1, a majority of respondents either agreed or strongly agreed that it was straightforward to comprehend and use. Regarding the assistance to developers in identifying ML-specific code smells, as demonstrated in Figure 3–Q2, the results indicate that the tool is indeed effective in helping to detect and address ML-specific code smell issues.

Respondents perceived the tool as dependable for detecting issues within Python-based ML applications (Figure 3–Q3). When asked about the tool's effectiveness in identifying ML-specific code smells, as presented in Figure 3–Q4, the majority responded with agree. This indicates that the tool is proficient at effectively highlighting identified code smells.

Shown in Figure 3–Q5, most of the respondents acknowledged that MLpylint facilitates a reduction in the time and effort required to identify code smells in ML applications. Moreover, they agreed that the code smells integrated were relevant (Figure 3–Q6) and commonly encountered in their daily ML development tasks, confirming the tool's applicability in the current industry landscape.

An open-ended question allowed participants to share feedback on their experience with the tool. Viewable in Table 9, they appreciated MLpylint for basic checks and saw a continued necessity for expert reviews. Suggestions on integrating it with existing tools like lint and extending support for more ML libraries were also noted. Respondents expressed concerns over dependency management and advocated for a more adaptable tool configuration.

Overall, we foresee two paths to go for better promote code smell analysis in the ML development ecosystem. Teams with mature DevOps pipelines may favor continuous execution in CI to prevent regressions in code quality. On the other hand, data-science notebooks could benefit from an on-demand extension that surfaces interactive advice only when explicitly invoked by the developer.

## 7 Threats and Validity

We discuss threats to the validity of this work following the classification proposed by Runeson *et al.* to comprehensively evaluate the robustness of the research findings [20].

**Reliability.** The reliability of this study is tied to several factors. First, ML-specific code smells are identified through a literature review. Various search engines and databases were utilized to contain as many studies as possible, ensuring a robust foundation for reliability. Second, the selection of methods and tools used in this study was grounded in another literature review and supplemented by consultations with experts in the field. This approach helped gather insights and knowledge, facilitating the creation of a reliable static code analysis tool with relevant development strategies.

In evaluating the tool's performance and ability to identify code smells, 160 open-source projects on GitHub were examined and used for manual testing to measure the tool's efficacy in real-world scenarios. Various data points and metrics were utilized to assess the tool's reliability comprehensively. Furthermore, feedback was sourced from a group of 15 experienced ML professionals from Ericsson to assess the tool's usefulness, ensuring a reliable and credible set of data for evaluating MLpylint's ability to identify code smells in Python ML applications. Despite these measures, the study has its limitations. MLpylint faced challenges in analyzing certain segments and occasionally produced false positives in the Code Smell Advice section, highlighting areas that require further refinement.

**Internal Validity.** The groundwork of this research was established through a literature review and the development of a set of criteria for identifying code smells, thereby, supporting this study's internal validity. Despite the strong foundation, the study recognizes the existence of factors that might pose threats. This includes the potential for unclear detection of code smells linked to external libraries and the somewhat narrow criteria for static analyzability. Also, the study could benefit from a more in-depth exploration of each code smell, potentially involving interviews with experts and industry leaders to uncover more detailed code smell patterns.

**External Validity.** This research leveraged a substantial analysis of numerous open-source projects, paired with validation through feedback from experienced professionals in the Telecom industry, to strengthen the applicability of MLpylint's functionalities across different contexts, thereby enhancing the study's external validity.

Nonetheless, we acknowledge that the study focuses on Python-based ML applications and relies heavily on external libraries. It is safe to say that these code smells are very much Python-related and would not apply to other programming languages.

**Construct Validity.** In the development of MLpylint, the literature was reviewed to identify, consider, and draw inspiration from similar existing tools. Moreover, expert insights were sought to better understand the utilization and design patterns used in these tools.

## 8 Conclusions and Future Work

This work aimed primarily to reduce technical debt and enhance the maintainability of ML software. We focused on investigating and mitigating ML specific code smells found in Python-based ML applications. Firstly, we sought to identify ML-specific code smells in such applications. Secondly, we aimed to create a static code analysis tool to pinpoint these specific smells. Thirdly, we evaluated the efficacy and performance of the developed tool. Lastly, feedback regarding the tool's feasibility and its value was collected from ML professionals at Ericsson.

The evaluation comprised the analysis of 160 open-source GitHub ML projects spanning over 10 million lines of code and detected 5,380 code smells in just 36 minutes. However, areas of improvement were recognized, especially concerning the precise identification of certain code smell patterns, which sometimes resulted in ambiguous identifications. We surveyed 15 ML professionals from Ericsson. The majority expressed positive views, emphasizing MLpylint's efficiency and its practical application in the industry. These professionals provided valuable suggestions for MLpylint's further development, including potential integration with other tools and the addition of code smells associated with more Python ML libraries. One initiative already in progress is the inclusion of MLpylint as a plugin to extend Pylint's capabilities to analyze ML code [10].

Looking forward, MLpylint has the potential for further development, with opportunities to broaden its analytical reach, increase its library support, and improve its accuracy in identifying code smells. We start with Zhang et al.'s smells as a first step, and we plan to include other smells in a future version of our tool. Furthermore, this research paves the way for future research to make further progress by utilizing MLpylint and helping to build a community where high ML software quality is standard. Furthermore, future research could explore other external Python libraries associated with machine learning. Different models could be examined to understand the prevalence of code smells, especially in large language models. This would entail identifying and analyzing new code smells, ensuring that the tool stays relevant and effective for Python ML development.

The focus could shift towards enhancing software engineering practices in ML application development. Future work can look into refining coding standards and integrating efficient workflows to streamline the development process, fostering a more productive and innovative developer environment. This involves exploring methods to easily integrate tools like MLpylint in the early stages of development to promote good coding practices from the outset.

Lastly, there is great value in expanding this research to include more open-source projects and delving deeper into evaluating each code smell checker. These approaches would provide a rich source of data and insight for the tool's future development. Therefore, future research should focus on adapting to the constantly changing field of ML application development through continuous validation.

## ACKNOWLEDGMENTS

We would like to thank the Ericsson's professionals for spending time on the MLpylint evaluation, giving us precious feedback.

## ARTIFACTS AVAILABILITY

The source code of the tool is currently publicly available at Github [9]. The list of repositories used in the evaluation is available in [11].

## REFERENCES

[1] Mohannad Alahdab and Gul Calikli. 2019. Empirical Analysis of Hidden Technical Debt Patterns in Machine Learning Software. *Lecture Notes in Computer Science* (11 2019), 195–202. https://doi.org/10.1007/978-3-030-35333-94
[2] Farhan Alebeisat, Zaid Alhalhouli, Tamara Alshabatat, and T.I. Alrawashdeh. 2018. Review of Literature on Software Quality. *8* 8 (2018), 32–42.
[3] Nicolás Cardozo. 2023. Prevalence of code smells in reinforcement learning projects. https://arxiv.org/abs/2303.10236
[4] Martin Fowler. 2002. *Refactoring: improving the design of existing code.* 256 pages. 31 https://doi.org/10.1007/3-540-45672-4\{\_
[5] J. Gesi, S. Liu, J. Li, I. Ahmed, N. Nagappan, D. Lo, E. S. De Almeida, P. S. Kochhar, and L. Bao. 2022. Code Smells in Machine Learning Systems. *ArXiv* (2022). https://arxiv.org/abs/2203.00803
[6] Hristina Gulabovska and Zoltán Porkoláb. 2019. Towards More Sophisticated Static Analysis Methods of Python Programs. In *2019 IEEE 15th International*

*Scientific Conference on Informatics*. 000225–000230. https://doi.org/10.1109/Informatics47936.2019.9119307

[7] Peter Hamfelt. 2023. MLpylint 0.0.3 – a static analysis tool for identifying ML-specific code smells. https://pypi.org/project/mlpylint/

[8] Peter Hamfelt. 2023. *MLpylint: Automating the Identification of Machine Learning-Specific Code Smells*. Master's thesis. Blekinge Institute of Technology.

[9] Peter Hamfelt. 2024. MLpylint – A static code analyzer tool for identifying ml-specific code smells. https://gitlab.com/peter.hamfelt/mlpylint

[10] Peter Hamfelt. 2024. Pylint plugin enhancing code analysis for machine learning and data science. https://github.com/pylint-dev/pylint-ml

[11] Peter Hamfelt, Ricardo Britto, Lincoln Rocha, and Camilo Almendra. 2025. Supplementary Material for the paper "Automatic Identification of Machine Learning-Specific Code Smells". https://doi.org/10.5281/zenodo.15793835.

[12] Peter Kokol. 2022. Software Quality: How Much Does It Matter? *Electronics* 11, 16 (2022), 2485. https://doi.org/10.3390/electronics11162485

[13] Guilherme Lacerda, Fabio Petrillo, Marcelo Soares Pimenta, and Yann Gaël Guéhéneuc. 2020. Code smells and refactoring: A tertiary systematic review of challenges and observations. *Journal of Systems and Software* 167 (9 2020), 110610. https://doi.org/10.1016/j.jss.2020.110610

[14] Sharon L. Lohr. 2021. *Sampling*. https://doi.org/10.1201/9780429298899

[15] Silverio Martínez-Fernández, Alberto Garcia-Perez, Xabier Larrucea, Miguel del Calvo-Flores, Pablo Garcia-Bringas, and Javier Tuya. 2022. Software Engineering for AI-Based Systems: A Survey. *ACM Transactions on Software Engineering and Methodology* 31, 2 (2022), 1–59. https://doi.org/10.1145/3487043

[16] Rana S. Menshawy, Ahmed H. Yousef, and Ashraf Salem. 2021. Code Smells and Detection Techniques: A Survey. In *2021 International Mobile, Intelligent, and Ubiquitous Computing Conference (MIUCC)*. 78–83. https://doi.org/10.1109/MIUCC52538.2021.9447669

[17] Edson Nascimento. 2020. Software engineering for artificial intelligence and machine learning software: A systematic literature review. *arXiv preprint arXiv:2011.03751* (2020). https://doi.org/10.48550/arXiv.2011.03751

[18] Philipp Offermann, Olga Levina, Marten Schönherr, and Udo Bub. 2009. Outline of a design science research process. In *Proceedings of the 4th International Conference on Design Science Research in Information Systems and Technology (DESRIST '09)*. ACM, New York, NY, USA, 1–11.

[19] Thanis Paiva, Amanda Damasceno, Eduardo Figueiredo, and Cláudio Sant'Anna. 2017. On the evaluation of code smells and detection tools. *Journal of Software Engineering Research and Development* 5, 1 (10 2017). https://doi.org/10.1186/s40411-017-0041-1

[20] Per Runeson and Martin Höst. 2009. Guidelines for Conducting and Reporting Case Study Research in Software Engineering. *Empirical Softw. Engg.* 14, 2 (apr 2009), 131–164. https://doi.org/10.1007/s10664-008-9102-8

[21] D. Sculley, Gary David Holt, Daniel Golovin, Eugene Davydov, Todd Phillips, Dietmar Ebner, Vinay Chaudhary, Michael Young, Jean-Francois Crespo, and D.H. Dennison. 2015. Hidden technical debt in Machine learning systems. *Neural Information Processing Systems* 28 (12 2015), 2503–2511. https://papers.nips.cc/paper/2015/file/86df7dcfd896fcaf2674f757a2463eba-Paper.pdf

[22] Andrew J. Simmons, Scott Barnett, Jessica Rivera-Villicana, Akshat Bajaj, and Rajesh Vasa. 2020. A large-scale comparative analysis of Coding Standard conformance in Open-Source Data Science projects. In *Proceedings of the 14th ACM / IEEE International Symposium on Empirical Software Engineering and Measurement (ESEM) (ESEM '20)*. Association for Computing Machinery, New York, NY, USA, 1–11. https://doi.org/10.1145/3382494.3410680

[23] Bart van Oort, Leonardo Cruz, Maurício Aniche, and Arie van Deursen. 2021. The Prevalence of Code Smells in Machine Learning projects. In *2021 IEEE/ACM 1st Workshop on AI Engineering - Software Engineering for AI (WAIN)* (Madrid, Spain). 1–8. https://doi.org/10.1109/WAIN52551.2021.00011

[24] Zhi Wan, Xin Xia, David Lo, and Gail C Murphy. 2021. How does Machine Learning Change Software Development Practices? *IEEE Transactions on Software Engineering* 47, 9 (Sept 2021), 1857–1871. https://doi.org/10.1109/TSE.2019.2937083

[25] Jing Wang, Raffi Khatchadourian, Mehdi Bagherzadeh, Rhia Singh, Ajani Stewart, and Anita Raja. 2021. An Empirical Study of Refactorings and Technical Debt in Machine Learning Systems. *International Conference on Software Engineering* (5 2021). https://doi.org/10.1109/icse43902.2021.00033

[26] Claes Wohlin, Per Runeson, Martin Höst, Magnus C. Ohlsson, Björn Regnell, and Anders Wesslén. 2012. *Experimentation in software engineering*. https://doi.org/10.1007/978-3-642-29044-2

[27] Hui Zhang, Luiz Angelo Da Silva Cruz, and Arie Van Deursen. 2022. Code Smells for Machine Learning Applications. https://doi.org/10.48550/arxiv.2203.13746 arXiv:2203.13746 [cs.SE]